# Modeling Language Variability

Hans Grönniger and Bernhard Rumpe

Software Engineering
RWTH Aachen University, Germany
`http://www.se-rwth.de`

**Abstract.** A systematic way of defining variants of a modeling language is useful for adapting the language to domain or project specific needs. Variants can be obtained by adapting the syntax or semantics of the language. In this paper, we take a formal approach to define modeling language variability and show how this helps to reason about language variants, models, and their semantics formally. We introduce the notion of *semantic language refinement* meaning that one semantics variant is implied by another. Leaving open all variation points that a modeling language offers yields the notion of the *inner semantics* of that language. Properties of the modeling language which do not depend on the selection of specific variants are called *invariant language properties* with respect to a variation point. These properties consequently follow from the inner semantics of a model or language.

## 1 Introduction

It has often been stressed that software is one of the most important drivers for innovation in many branches of industry. Developers are faced with the challenge to produce high quality, increasingly complex solutions in a short period of time.

Model-based software development is regarded as one instrument to cope with the challenges. Standard modeling languages like UML [OMG09] or domain specific languages (DLSs) are employed to increase the level of abstraction and automation while at the same time lowering the complexity. Especially in the context of robust, reliable systems development, the modeling languages used have to be defined precisely to allow for rigorous analysis of models and correct code generation.

The precise definition of a modeling language involves syntax and semantics [HR04]. Formal semantics is advantageous because it helps to avoid misunderstandings between people and may enable interoperability between tools. But even if a formal modeling language exists, a new class of systems like highly robust and reliable systems or a specific application domain may require adaptation of the language. A language may be changed to incorporate new language constructs, to disallow others for methodological or safety reasons, or to be semantically adjusted to a specific platform. This variability of a modeling language is subject of the paper.

We provide a formal account on language variability based on our classification in [CGR09]. On the one hand, the formalization brings light into how a language can be adapted to specific requirements. On the other hand, it serves as a basis to define language variants formally. This allows us to reason about language (especially semantic) variants. The theoretical work is also equipped with tool support. Complete language definitions including all aspects of syntax and semantics and their variants are handled using the tools MontiCore [KRV08] and Isabelle/HOL [NPW02]. MontiCore is a framework for the development of modular (domain-specific) modeling languages while Isabelle/HOL is a theorem prover with higher-order logic and suitable to encode various language aspects.

The paper is structured as follows. The basic constituents (syntax, semantics) of a modeling language that may be subject to variability are introduced in Section 2. In Section 3, a formal characterization of language variants and a method to define variants is presented. As an example application, we outline how semantic variants can be compared formally in Section 4. In this section, we also introduce the concepts of semantic language refinement, inner semantics, and invariant language properties. Section 5 sketches the available tool support. In Section 6, we discuss related work. Section 7 concludes the paper.

## 2   Language Constituents

A precise definition of a modeling language consists of the following elements, see also [HR04, CGR09].

*Concrete Syntax.* The concrete syntax is the representation of the model with which a user interacts. This may be a graphical or textual notation or a mixture of both. We denote the set of all models of a modeling language in concrete syntax by $\mathcal{CS}$.

*Abstract Syntax.* The abstract syntax represents the structural essence of a language [Wil97]. For a textual syntax this may be given as abstract syntax trees generated by a parser. In case of graphical models, metamodels (e.g., defined in MOF [OMG06a]) are typically used. The set of all models of a modeling language in abstract syntax is denoted by $\mathcal{AS}$.

Additionally, a set of well-formedness rules or context conditions is defined to rule out certain models based on syntactic criteria. A typical example is that, in an automaton language, sources and targets of transitions have to exist so there are no dangling start or end points. But also the question of whether a model, e.g., a class diagram containing OCL constraints, is well-typed is addressed on the syntactical level. We assume a predicate

$$\text{wellformed} : \mathcal{AS} \to bool$$

to decide if a model is well-formed. The set of all well-formed models $\mathcal{AS}^{\text{wf}}$ of a language hence is

$$\mathcal{AS}^{\text{wf}} = \{m \in \mathcal{AS} \,|\, \text{wellformed}\, m\}$$

We may also define additional constraints that rule out models for methodological or safety reasons, potentially restricting the expressiveness of the language. A more detailed explanation will be given in the next section.

A model in concrete syntax is associated with (or mapped to) a model in abstract syntax. Since typically not all models from $\mathcal{CS}$ are well-formed, parsing is a partial mapping from concrete to abstract syntax:

$$\mathrm{p} : \mathcal{CS} \rightharpoonup \mathcal{AS}^{\mathrm{wf}}$$

*Reduced Abstract Syntax.* It is often advisable to reduce the number of language constructs for a simplification of semantic considerations. This is possible for each language construct that can be expressed by others (such as $\forall$ by $\neg\exists\neg$ in predicate logic). This reduces set $\mathcal{AS}^{\mathrm{wf}}$ to a subset $\mathcal{AS}^{\mathrm{red}} \subseteq \mathcal{AS}^{\mathrm{wf}}$ with a syntactic transformation $\mathrm{t}$ to convert models into the reduced abstract syntax, i.e.,

$$\mathrm{t} : \mathcal{AS}^{\mathrm{wf}} \rightarrow \mathcal{AS}^{\mathrm{red}} \text{ with } \mathcal{AS}^{\mathrm{wf}} \supseteq \mathcal{AS}^{\mathrm{red}}$$

*Semantic Domain.* By mapping models to elements of a semantic domain $\mathcal{S}$, the models obtain their meanings. The semantic domain is required to be well-known and understood and it should be based on a well-defined mathematical theory.

Our approach to semantics uses the system model [BCGR09a, BCGR09b] which characterizes the structure, behavior, and interaction of objects in object-based systems. Thus, our focus is on semantics of object-based modeling languages. However, the variability mechanisms still apply if another semantic domain is used. The system model definitions are built on simple mathematical concepts like sets, relations, and functions. It is important to note that one element in the system model represents a single, complete object-based system. This means that the meaning of a model is directly represented as properties of possible implementations. The system model is underspecified to allow, for example, freedom of implementation when mapping a model to executable code.

For later reference, we introduce but a few system model concepts. Generally, elements of object-based systems are introduced as elements of underspecified universes leaving open the exact structure or number of elements. There is, for each system $s \in$ SystemModel, a set of class names (or just classes, for short) UCLASS$_s$. In the following, we leave out the index $s$ but a specific system is assumed implicitly if not stated otherwise. A class $C_1$ may be in a subset relation to a class $C_2$ which is denoted as $(C_1, C_2) \in$ sub $\subseteq$ UCLASS $\times$ UCLASS. There is also a set of operation names (method signatures) UOPN. With function classOf : UOPN $\rightarrow$ UCLASS the defining class for an operation is obtained. Function nameOf determines the name of the operation, function params yields the set of all possible parameter assignments and function resType gives the return type of an operation. Types are elements of a universe UTYPE and there is a carrier set of values from universe UVAL associated with each type: CAR : UTYPE $\rightarrow \wp(\mathrm{UVAL})$. $\wp(X)$ denotes the set of all subsets of $X$ (power set).

*Semantic Mapping.* The semantic mapping sem finally relates models of the reduced abstract syntax to elements of the semantic domain. Characteristic of our loose approach is a set-valued or predicative semantic mapping of the form

$$\mathrm{sem} : \mathcal{AS}^{\mathrm{red}} \rightarrow \wp(\mathcal{S})$$

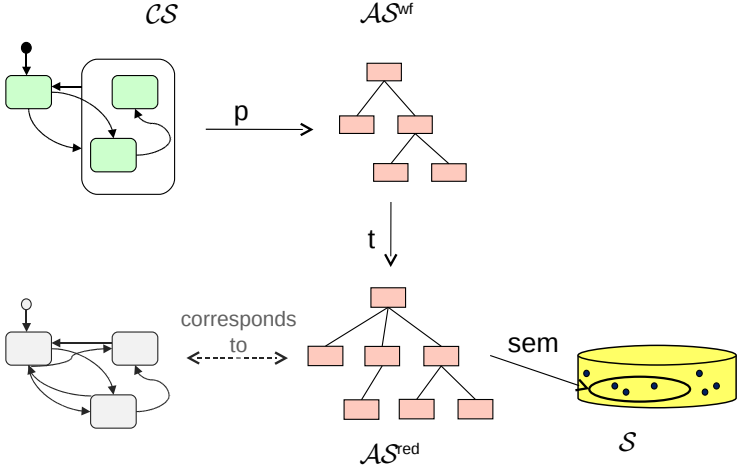

**Fig. 1.** From syntax to semantics for a Statechart model

The semantics of a model $m$ is therefore the set $sem(m)$ of elements in the domain $\mathcal{S}$. If the system model is used for $S$, then the model's meaning is the set of all possible realizations.

Using the system model as a single semantic domain and the set-valued semantic mapping enable a straightforward treatment of composition and refinement of possibly incomplete and underspecified models of various modeling languages [Rum96]. For example, the integrated semantics of models $m_1, \ldots, m_n$ from possibly different languages is given as

$$sem_1(m_1) \cap \ldots \cap sem_n(m_n)$$

In the same way, a model $m'$ is a refinement of model $m$, exactly if

$$sem(m') \subseteq sem(m)$$

The whole chain from syntax to semantics is illustrated in Fig. 1. The example shows a hierarchical automaton (Statechart) in concrete syntax. Its abstract syntax is transformed into a conceptually reduced abstract syntax. For example, the automaton is flattened and the concept of hierarchy can be eliminated in the abstract syntax. Note that the (abstract) syntax of the resulting automaton will be more verbose compared to the original version. With the help of the semantic mapping, the automaton is mapped into the system model. Its semantics is given as a set of systems in the system model. These systems have to obey the properties introduced by the model. Hence, in the semantic mapping, we have to define ways to associate Statechart states with concepts in the system model (like classes, attributes, etc.). Additionally, we need means to encode the enabledness of transitions and their effect when actually executed.

# 3     Language Variants

A modeling language should be defined precisely but should not be completely fixed. Sustaining a certain degree of flexibility regarding a language's syntax or semantics allows for adapting it to project or domain specific needs, or to enable modeling of new classes of systems. This idea has also been incorporated in the definition of UML where the informal semantics is equipped with semantic variation points subject to specific interpretation. At present, the UML standard itself regards semantic variation points as "less precisely defined dimensions" [OMG09]. We take a formal approach to define the possible variability in a language definition thereby substantiating our classification in [CGR09]. Afterwards, we present an intuitive way to document language variants.

## 3.1     Classification of Language Variability

In the previous section, we defined the constituents of a modeling language and their relations. A model in concrete syntax is translated into its abstract representation which then is (optionally) transformed into a conceptually simplified version. Based on this, the semantic mapping associates sets of elements of the semantic domain with the model. To summarize, we have the sequence

$$\mathcal{CS} \overset{\mathrm{p}}{\to} \mathcal{AS}^{\mathrm{wf}} \overset{\mathrm{t}}{\to} \mathcal{AS}^{\mathrm{red}} \overset{\mathrm{sem}}{\to} \wp(S)$$

In this section, we discuss means to define variants of a modeling language by adapting one or more elements of the above sequence.

*Presentation Variability.* A modeling language may offer *presentation options*, a term also coined in the UML standard. Presentation options allow for representing models differently in concrete syntax without changing a model's abstract syntax. Formally, a language contains presentation options, if

$$\exists m_1, m_2 \in \mathcal{CS} : m_1 \neq m_2 \wedge \mathrm{p}(m_1) = \mathrm{p}(m_2)$$

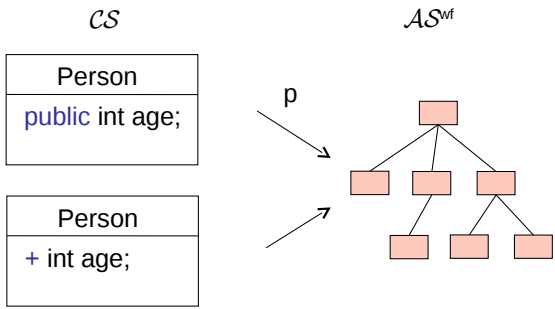

**Fig. 2.** Presentation option: Modifier representation in a class diagram

As shown in Fig. 2, for example, we have different ways to represent a public class modifier in UML: We can use the keyword `public` but equivalently the symbol `+`. The resulting abstract syntax, however, stays the same. Variants of presentation options result in changes of $\mathcal{CS}$ and p, say $\mathcal{CS}_v$ and $p_v$, by introducing, eliminating or changing existing ones. Models contained in both variants still have the same abstract syntax:

$$\forall m \in \mathcal{CS}_v \cap \mathcal{CS} : p_v(m) = p(m)$$

Additionally, every model can be expressed without choosing the presentation option variant:

$$\forall m_1 \in \mathrm{dom}(p_v) : \exists m_2 \in \mathrm{dom}(p) : p_v(m_1) = p(m_2)$$

Another form of presentation variability is what we call *abbreviations* or *extended constructs*: The syntax may contain certain constructs that help to enhance readability and comfort but which can be eliminated by some syntactic transformation t without loosing expressiveness of the language. All models which do not use extended constructs remain identical under t, i.e.,

$$\forall m \in \mathcal{AS}^{\mathrm{red}} : t(m) = m$$

The models that actually get transformed are contained in $\mathcal{AS}^{\mathrm{wf}} \backslash \mathcal{AS}^{\mathrm{red}}$. Variability in abbreviations means adapting $\mathcal{AS}^{\mathrm{wf}}$ and t, to $\mathcal{AS}^{\mathrm{wf}}_v$ and $t_v$ say. Consider, for example, a reduced abstract syntax for Statecharts $\mathcal{AS}^{\mathrm{red}}$ which contains flat automata only (see Fig. 1). Hierarchy can be added to or removed from Statecharts without changing expressiveness [Rum04], but we obtain a larger set of expressible models when adding hierarchy, i.e., $\mathcal{AS}^{\mathrm{wf}}_v \supseteq \mathcal{AS}^{\mathrm{red}}$. Models that do not contain an extended construct variant (e.g., hierarchy) are transformed equally under $t_v$:

$$\forall m \in \mathrm{dom}(t_v) \cap \mathrm{dom}(t) : t_v(m) = t(m)$$

And we can still represent each model without the abbreviation:

$$\forall m_1 \in \mathrm{dom}(t_v) : \exists m_2 \in \mathrm{dom}(t) : t_v(m_1) = t(m_2)$$

As abbreviations do not show up in the reduced abstract syntax, semantics of these constructs is defined in two steps, the first one being the transformation to $\mathcal{AS}^{\mathrm{red}}$ for which semantics is defined via the semantic mapping sem. Summarizing, variants of presentation options have an effect on the concrete syntax. Variants of abbreviations have an effect on the full abstract syntax. Both do not change the reduced abstract syntax and are called presentation variability.

*Syntactic Variability.* We now consider language variants that also have an impact on the reduced abstract syntax $\mathcal{AS}^{\mathrm{red}}$. The syntax of a language may allow the use of *stereotypes*. A set of defined stereotypes (e.g., as part of a profile in case of UML) is a syntactic variant of the language. We assume a function variant allowedStereotypes$_v$ that checks if only the chosen stereotypes are used, i.e.,

$$\mathcal{AS}_v^{\mathrm{red}} = \{m \in \mathcal{AS}^{\mathrm{red}} \,|\, \mathrm{allowedStereotypes}_v(m)\}$$

An example for this kind of variability are priorities of transitions in a Statechart. A stereotype `<<prio:outer>>` attached to a hierarchical Statechart model would override the default priority rule that the innermost enabled transition is taken if there are multiple transitions with the same trigger enabled in the same step, see also [Rum04].

Another form of syntactic variability is given by so called *language parameters*, also termed language embedding in [KRV08]. Consider again, for example, the language of Statecharts in which transitions may be guarded by a precondition. The language in which this condition is expressed is not specified. A natural candiate language would be OCL [OMG06b] but we may allow any other constraint language or a variant thereof that is suitable for the intended application. Hence, a syntax can be equipped with parameters $\mathcal{AS}^{\mathrm{red}}(p_1, \ldots, p_n)$. Variants can then be specified by assigning concrete languages to the parameters $p_1, \ldots, p_n$.

As a last form of syntactic variability, we consider general *language constraints*. A language is further constrained to disallow certain models syntactically. It may be the case that this results in a less expressive language. Formally, a variant $\mathcal{AS}_v^{\mathrm{red}}$ is given by models which fulfill further constraints stated, for example, in the predicate constr$_v$:

$$\mathcal{AS}_v^{\mathrm{red}} = \{m \in \mathcal{AS}^{\mathrm{red}} \,|\, \mathrm{constr}_v(m)\}$$

The expressiveness of the language is preserved if

$$\forall m_1 \in \mathcal{AS}_v^{\mathrm{red}} : \exists m_2 \in \mathcal{AS}^{\mathrm{red}} : \mathrm{sem}(m_1) = \mathrm{sem}(m_2)$$

It is, for example, the goal of modeling or programming guidelines [Mat07, MIS] to restrict the use of certain (e.g., unsafe) language constructs while preserving the expressiveness. Restricting the expressiveness might be useful in situations in which a target platform may not be powerful enough to implement the models.

*Semantic Variability.* While UML only uses the term semantic variation point, we further subdivide semantic variability into *semantic mapping variability* and *semantic domain variability*. A helpful analogy might be to see the variability of the semantic mapping similar to configuration options of a code generator while variability of the semantic domain has its analogy with properties of an underlying run-time system or target platform.

By selecting variants for the semantic domain $\mathcal{S}$, we obtain an adapted domain $\mathcal{S}_v$ in which elements have certain additional properties, for example, encoded in a predicate prop$_v$:

$$\mathcal{S}_v = \{s \in \mathcal{S} \,|\, \mathrm{prop}_v(s)\}$$

Regarding semantic domain variability, the system model already contains explicit variability in form of extensions through optional definitions. It provides, for instance, different notions of type-safe method overriding or optional constraints to allow single inheritance only.

As an example for semantic domain variability, we show two variants for type-safe overriding of operations in a subclass. The first variant contains the well-known formalization of co-variant extension of parameters and contra-variant restriction of return values for operations in the subclass [Mey97]. In the system model, this can be expressed by a subset relation on the sets of all possible parameters and all possible return values, respectively:

$$\forall op_1 \in \mathrm{UOPN}, c \in \mathrm{UCLASS} : c \,\mathrm{sub}\,\mathrm{classOf}(op_1) \implies$$
$$\exists op_2 \in \mathrm{UOPN} : \mathrm{classOf}(op_2) = c \,\wedge$$
$$\mathrm{nameOf}(op_1) = \mathrm{nameOf}(op_2) \,\wedge$$
$$\mathrm{params}(op_1) \subseteq \mathrm{params}(op_2) \,\wedge$$
$$\mathrm{CAR}(\mathrm{resType}(op_1)) \supseteq \mathrm{CAR}(\mathrm{resType}(op_2))$$

The second variant is stricter as it does not allow a modification of the operation's signature in terms of possible values for parameters and the return type:

$$\forall op_1 \in \mathrm{UOPN}, c \in \mathrm{UCLASS} : c \,\mathrm{sub}\,\mathrm{classOf}(op_1) \implies$$
$$\exists op_2 \in \mathrm{UOPN} : \mathrm{classOf}(op_2) = c \,\wedge$$
$$\mathrm{nameOf}(op_1) = \mathrm{nameOf}(op_2) \,\wedge$$
$$\mathrm{params}(op_1) = \mathrm{params}(op_2) \,\wedge$$
$$\mathrm{CAR}(\mathrm{resType}(op_1)) = \mathrm{CAR}(\mathrm{resType}(op_2))$$

Variants of a semantic mapping arise as alternative definitions of (parts of) the semantic mapping, for example

$$\mathrm{sem}_{v1}, \mathrm{sem}_{v2} : \mathcal{AS}^{\mathrm{red}} \to \wp(\mathcal{S})$$

Considering a Statecharts semantics again, a mapping variant could be the different choices of representing Statecharts states (syntax) as, for example, a simple enumeration in a class or using the state pattern [GHJV95].

Note that semantic variability is transparent to the modeler. But it may be necessary to allow the modeler to select one or the other interpretation of a construct. We propose to model these interpretation choices as syntactic variability by providing corresponding stereotypes. A modeler can then select the semantics of certain constructs by using appropriate stereotypes. With this approach, we transfer semantic variation points to syntactic ones.

## 3.2   Documentation of Language Variability

We propose to model variation points and variants in a language by feature diagrams [CE00]. Fig. 3 contains a feature diagram representing a generic structure to model variants of a language L. We do not show concrete variants which depend on a specific language and which would be inserted under the corresponding nodes.

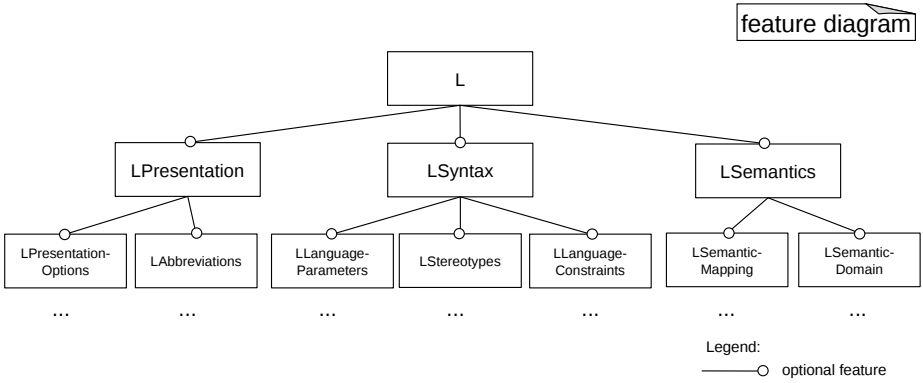

**Fig. 3.** Template to document variability of a language L

A supplement description of the variability can be given to explain their raison d'être and to point to formal definitions of the variants or other documentation. Since our main focus currently is on UML-like modeling languages, we refrained from expressing the variability in UML itself. This would certainly be possible but might be more confusing since UML models would be used also on the language definition level.

In Fig. 4 we present an excerpt of the feature diagram of a Statechart language containing the variants discussed previously in this paper. The choice of a feature is always optional but there may be exclusive alternatives. It is, for example, not reasonable to choose both mappings for the decision of which transitions are enabled. Likewise we can only use one specific interpretation of Statechart states. There is an inclusion constraint for priority-based selection of transitions since priorities refer to syntactical as well as semantic variants at the same time (encode the priorities as stereotypes and select a semantic mapping that can handle them).

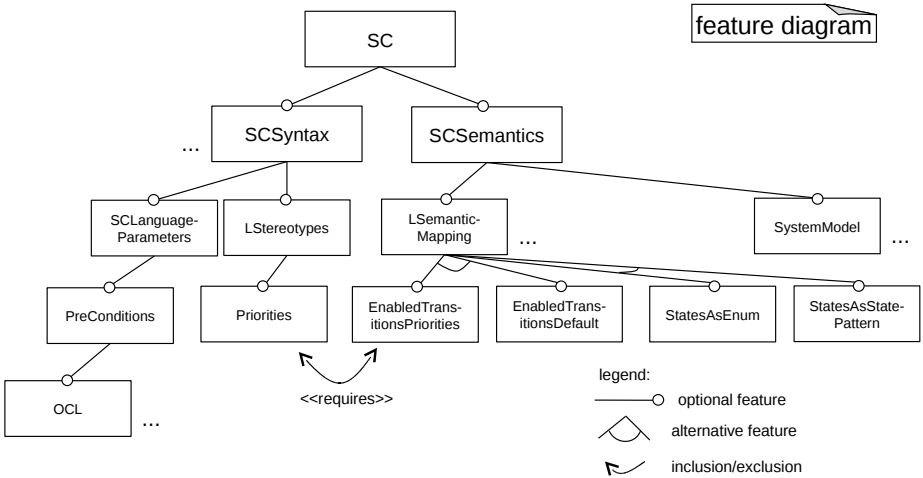

**Fig. 4.** Feature diagram

## 4   Comparison of Semantic Variants

Using our formal notion of language variants, it is possible to compare language variants formally and derive properties of the (relationship between) variants. Consider two semantic variants of the same language, e.g.,

$$\mathrm{sem}_{v1} : \mathcal{AS}^{\mathrm{red}} \to \wp(\mathcal{S}_{v1})$$
$$\mathrm{sem}_{v2} : \mathcal{AS}^{\mathrm{red}} \to \wp(\mathcal{S}_{v2})$$

An interesting property is if variant $v2$ is a *semantic language refinement* of the semantic variant $v1$. Note that we discuss language refinement here and do not talk about refinement of models or the modeled system.

We define that language variant $v2$ is a semantic language refinement of variant $v1$ exactly if for all models the sets generated by the respective semantic mapping are in a subset relation, i.e.,

$$\forall m \in \mathcal{AS}^{\mathrm{red}} : \mathrm{sem}_{v1}(m) \supseteq \mathrm{sem}_{v2}(m)$$

This implies that all properties $\phi$ of a model $m$ which hold in variant $v1$ are preserved in variant $v2$:

$$\forall s \in \mathrm{sem}_{v1}(m) : \phi(s) \implies \forall s \in \mathrm{sem}_{v2}(m) : \phi(s)$$

Semantic language refinement is an important property if we consider for example tool integration. Assume that one tool for formal analysis uses (and correctly implements) language variant $v2$. Another tool for code generation correctly implements variant $v1$. If we show that variant $v2$ is a language refinement of $v1$ then we can be sure that analysis results obtained by the analysis tool are preserved in the second tool for code generation.

Let $\mathrm{sem}_{v1}$ be the Statechart semantics where the realization of states (either as an enumeration in a class or using the state pattern) is left open. Obviously, a semantic variant in which one of the alternatives is selected is always a subset of $\mathrm{sem}_{v1}$ for any model. A property $\phi$ which holds for $\mathrm{sem}_{v1}$ hence also holds under $\mathrm{sem}_{v2}$. Since we did not select a specific variant in $\mathrm{sem}_{v1}$, we say that $\phi$ is an *invariant property* with respect to the variation point on the interpretation of Statechart states. The property may be globally invariant (valid for all models) or locally invariant (for at least a single model). Not choosing any specific variant for any semantic variation point yields the notion of *inner semantics* of a modeling language. Properties shown for the inner semantics are intrinsic language properties and are agnostic to variant selection.

# 5  Tool Support

We have developed tool support in order to a) specify a machine-readable, checkable semantics that can directly be used for verification purposes, and b) to better control and quality check the different artifacts by using standard tools, e.g., version control. Fig. 5 gives an overview of the approach when defining the semantics of a language with tool support. First, the (domain specific) modeling language concepts are specified using a MontiCore grammar. MontiCore [KRV08] is a framework for the textual definition of languages based on an extended context-free grammar format. This format enables a modular development of the syntax of a language by providing modularity concepts like language inheritance and language parameters/embedding. MontiCore has an integrated, consistent definition of concrete and abstract syntax which also provides meta-modeling concepts like associations and inheritance [KRV07]. Framework functionality helps developers also to define well-formedness rules and, for example, the implementation of generators. To provide the semantics developer with maximum flexibility but also with some machine-checking (i.e., type checking) of the semantics and the potential for real verification applications, we use the theorem prover Isabelle/HOL for

- the formalization of the system model as a hierarchy of theories, including its semantic domain variants. This step has to be done once. All following language definitions which should be based on the system model can re-use this effort.
- the representation of the abstract syntax of the language as a deep embedding, including its syntactic variants. The translation of a MontiCore grammar to Isabelle/HOL abstract syntax data types is automated. Only manual configuration of variants is needed. We decided to follow a deep embedding approach (explicit representation of the abstract syntax in Isabelle/HOL) because this allows us also to reason about syntactical entities and the semantic mapping. With a shallow embedding (encoding properties of systems of the system model directly for a given model) this would not be possible.
- the semantic mapping which maps the generated abstract syntax to predicates over systems of the formalized system model, including its semantic mapping variants.
- specification of context conditions that may be helpful when doing verification on well-formed models, including variants in context conditions.

Details of the approach can also be found in [GRR09] and [Grö10]. We give a small example in which we prove in Isabelle/HOL that the stronger variant for type-safe operations refines the weaker variant. For this, we first need the encoding of the two predicates in Isabelle/HOL. The first one is given in Fig. 6. The figure shows an excerpt of an Isabelle/HOL theory which defines the predicate `valid-TypeSafeOps` as a function for systems in the system model (l. 1). In general, each definition in Isabelle/HOL is parameterized with the system model. As with the index $s$ used earlier this means that the predicate is valid or not for a given system `sm`. For example, `UCLASS sm` (l. 4) is the specific universe

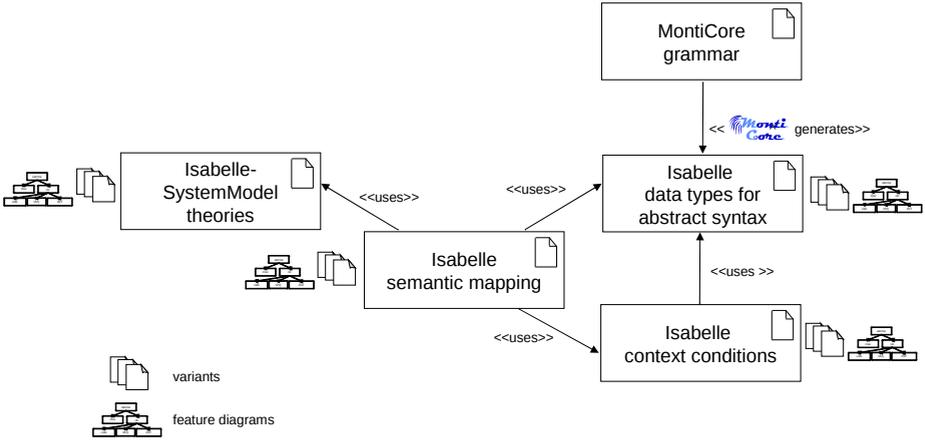

**Fig. 5.** Overview of approach to define a modeling language including its variants

```
                          TypeSafeOps
1    fun valid-TypeSafeOps :: "SystemModel ⇒ bool"
2    where
3     "valid-TypeSafeOps sm = (
4        ∀ op1 ∈ UOPN sm . ∀ C ∈ UCLASS sm .
5          (sub sm C (classOf sm op1)) ⟶
6            (∃ op2 ∈ UOPN sm .
7              classOf sm op2 = C ∧
8              nameOf op1 = nameOf op2 ∧
9              params sm op1 ⊆ params sm op2 ∧
10             CAR sm (resType sm op2) ⊆ CAR sm (resType sm op1)
11           )
12     )"
```

**Fig. 6.** Part of a theory of the system model in Isabelle/HOL encoding the predicate of type-safe overriding of operations

of class names of the system `sm`. Similarly, other universes and functions are parameterized with the concrete system `sm`. Apart from slight notational differences, the predicate is a direct translation of the predicate given in Sect. 3.1. The predicate for the stronger variant is similar, only the subset relation is replaced by equality. We do not give the whole definition but the predicate is called `valid-TypeSafeOpsStrict`.

To prove the refinement from `TypeSafeOps` to `TypeSafeOpsStrict` we have to show that the set of systems with the second property is a subset of the set of systems with the first property. The actual proof is now given in Fig. 7. In Isabelle/HOL, this is a lemma which is given the name `TypeSafeOpsImplStrict` (l. 1). Applying the predicate definitions (ll. 4,5) the proof can be finished automatically by Isabelle (l. 6).

```
                          ┌── RefinedTypeSafeOps ──────────────────────────────┐
 1    lemma TypeSafeOpsImplStrict :
 2        "{sm | sm . valid-TypeSafeOpsStrict sm}
 3                ⊆ {sm | sm . valid-TypeSafeOps sm}"
 4    apply (unfold valid-TypeSafeOps.simps)
 5    apply (unfold valid-TypeSafeOpsStrict.simps)
 6    by best
```

**Fig. 7.** Part of a theory containing the proof that the strict variant refines the weaker variant

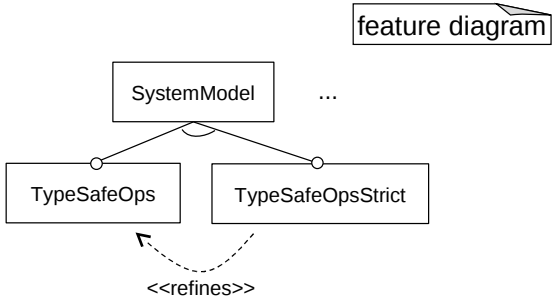

**Fig. 8.** Documentation of refinement relationship between variants in the feature diagram

Having shown that a variant refines another, we propose to update the feature diagram accordingly. Fig. 8 contains an excerpt of the feature diagram of the system model containing the information that `TypeSafeOpsStrict` is a semantic domain refinement of variant `TypeSafeOps`. As before, we assume that additional information, documentation may be attached to the feature diagram. This could be a link to the actual Isabelle/HOL proof that establishes the refinement relation. Other examples using the tool-supported approach can be found in [Grö10] and [GRR09].

## 6   Related Work

Presentation and semantic variants are also covered informally in the UML standard [OMG09]. We state precisely what kinds of variability may be found in a modeling language and document variants using feature diagrams.

Feature diagrams are also used in [Völ08] to define a family of architecture description languages. Formal semantics is not addressed. In the area of semantics, semantic variability is covered to some extent.

Template semantics [NAD03] as well as templatable metamodels [CMTG07] can be used to describe semantics with variation points. None of the mentioned work discusses the possibility to compare language variants. [TA06] examines different variants of formal Statecharts semantics but does not address formal relationships between the variants.

Informal comparisons of Statecharts variants can, for example, be found in [Bee94, CD07].

Tool support to define modeling languages, including their formal semantics, is for example described in [CSAJ05]. This work presents semantic anchoring which means to transform the abstract syntax of a language into the abstract syntax of a language with known, formal semantics, for example Abstract State Machines (ASMs). [KM08] contains an Alloy-based approach that also allows to handle complete language definitions - from syntax, well-formedness of models to operational semantics. Mainly focusing on operational semantics these approaches have problems with underspecification and are not capable of integrating multiple languages into one common semantic domain easily.

## 7   Conclusion

We have formally described the constituents of a modeling language and how they can be varied to obtain modeling language variants. As an example application of precise modeling language variants, we have introduced the notion of semantic language refinement. Given two semantics variants of a language this notion defines if it is safe to use the one instead of the other variant. Additionally, we introduced the concept of inner semantics of a language, meaning to leave open all available variation points, and the notion of invariant properties with respect to a variation point. We have furthermore sketched the available tool support for complete language definitions with variability and how it can be applied to verify relationships between semantic variants.

Future work is concerned with investigating other relationships between language variants. Additionally, this work needs to be applied to, for example, the UML, or to various domain specific languages and needs to be explored in practice.